\documentclass[aps,pra,a4paper,twocolumn,floatfix]{revtex4}
\usepackage[latin1]{inputenc}
\usepackage{amsfonts}
\usepackage{amssymb}
\usepackage{amsmath}
\usepackage{calc}
\usepackage{graphicx}
\usepackage{epsfig}
\usepackage{bm}
\usepackage{ulem}
\usepackage{setspace}
\usepackage{subfigure}

\def\opone{\leavevmode\hbox{\small1\kern-3.8pt\normalsize1}}

\begin{document}

\title{On the impossibility of faithfully storing single-photons \\ with the three-pulse photon echo}
\date{\today}
\author{Nicolas Sangouard$^{1},$ Christoph Simon$^{2},$ Ji\v{r}\'{i} Min\'a\v{r}$^{1},$ Mikael Afzelius$^{1},$ Thierry Chanelière$^{3},$ Nicolas Gisin$^{1},$ Jean-Louis Le Gou$\ddot{\text{e}}$t$^{3},$ Hugues de Riedmatten$^{1},$ and Wolfgang Tittel$^{2}.$}
\affiliation{$^{1}$Group of Applied Physics, University of Geneva, Switzerland \\
$^{2}$Institute for Quantum Information Science and Department of Physics and Astronomy, University of Calgary, Calgary, Alberta T2N 1N4, Canada\\
$^{3}$Laboratoire Aimé Cotton, CNRS-UPR 3321, Université Paris-Sud, 91405 Orsay cedex, France}

\begin{abstract}
The three-pulse photon echo is a well-known technique to store intense light pulses in an inhomogeneously broadened atomic ensemble. This protocol is attractive because it is relatively simple and it is well suited for the storage of multiple temporal modes. Furthermore, it offers very long storage times, greater than the phase relaxation time. Here, we consider the three-pulse photon echo in both two- and three-level systems as a potential technique for the storage of light at the single-photon level. By explicit calculations, we show that the ratio between the echo signal corresponding to a single-photon input and the noise is smaller than one. This severely limits the achievable fidelity of the quantum state storage, making the three-pulse photon echo unsuitable for single-photon quantum memory.
\end{abstract}

\maketitle
\section{Introduction}
\textit{Motivation}\\
The distribution of entanglement over long distances is at the heart of future quantum networks. It may rely on quantum repeaters, which require photonic quantum memories enabling the storage of a large number of temporal modes for very long times \cite{Sangouard09}. Light storage based on photon-echo techniques in inhomogeneously broadened atomic ensembles has been widely studied in the classical regime \cite{Mossberg82}. These techniques are naturally suited to store intense light pulses in multiple temporal modes. Storage and retrieval of up to 1760-pulse sequences has been demonstrated \cite{Iin95}. Furthermore, when the photon-echo is generated with a specific sequence of three pulses, the so-called three-pulse photon echo, it further offers very long storage times, greater than the phase relaxation time, limited by the population relaxation time only \cite{Ohlsson03}. This technique could thus be very attractive, a priori, in the context of quantum repeaters, and it is natural to wonder whether it can be extended to the storage of single-photons.\\

\textit{Result}\\
It has been shown already that the simplest form of the photon echo, namely the two-pulse photon echo, is not suited for quantum storage \cite{Ruggiero09}. However, a recent paper \cite{Ham09} might let us believe, at first sight, that the photon echo in its three-pulse version could be used for the storage of light at the single-photon level. We thus analyze this last technique in detail and we show in this paper, that it does not preserve the fidelity of the stored quantum state. The main reason is that the rephasing process, induced by intense pulses and leading to the collective emission of an echo, transfers a large number of atoms to an excited state. The excitation of the medium is followed by spontaneous emission of many photons in random spectral, temporal, spatial and polarization modes, including those that the stored photon may originally have been prepared in. This makes complete filtering of fluorescence impossible, and strongly degrades the fidelity of the echo signal associated with a few-photon input. Furthermore, the rephasing process is only partial, which intrinsically limits the storage efficiency. These fundamental limitations, which also apply to the protocol of Ref. \cite{Ham09}, make it impossible to implement a high-fidelity and high-efficiency quantum memory based on the three-pulse photon echo in the few-photon regime.\\

\textit{Methods}\\
To evaluate the response of an inhomogeneously broadened atomic ensemble to a sequence of many resonant pulses, we follow the semi-classical approach that originated in Ref. \cite{Abella66, Mossberg79}. First, we focus on the temporal evolution of a single atom. Then, we sum over all individual responses to get (i) the macroscopic polarization of the sample and thus the intensity of the echo (ii) the average population on the excited state and thus the noise. We then deduce the signal-to-noise ratio for the specific case of single-photon storage. This approach also allows to bound the storage efficiency. However, its does not give any information about additional limitations caused by the propagation of the light fields, which could be obtained by solving Maxwell-Bloch equations. Yet, it is interesting to note that a model as simple as the one based on the dynamics of a single atom clearly shows the impossibility to faithfully store individual photons with the three-pulse photon echo.  \\

\textit{Outline}\\
In the next section, we present our analysis of the three-pulse photon echo in two-level systems. In section \ref{section3}, we extend our study to three-level systems. The last section is devoted to our conclusion.

\section{Two-level systems}
\begin{figure}
{\includegraphics[scale=0.25]{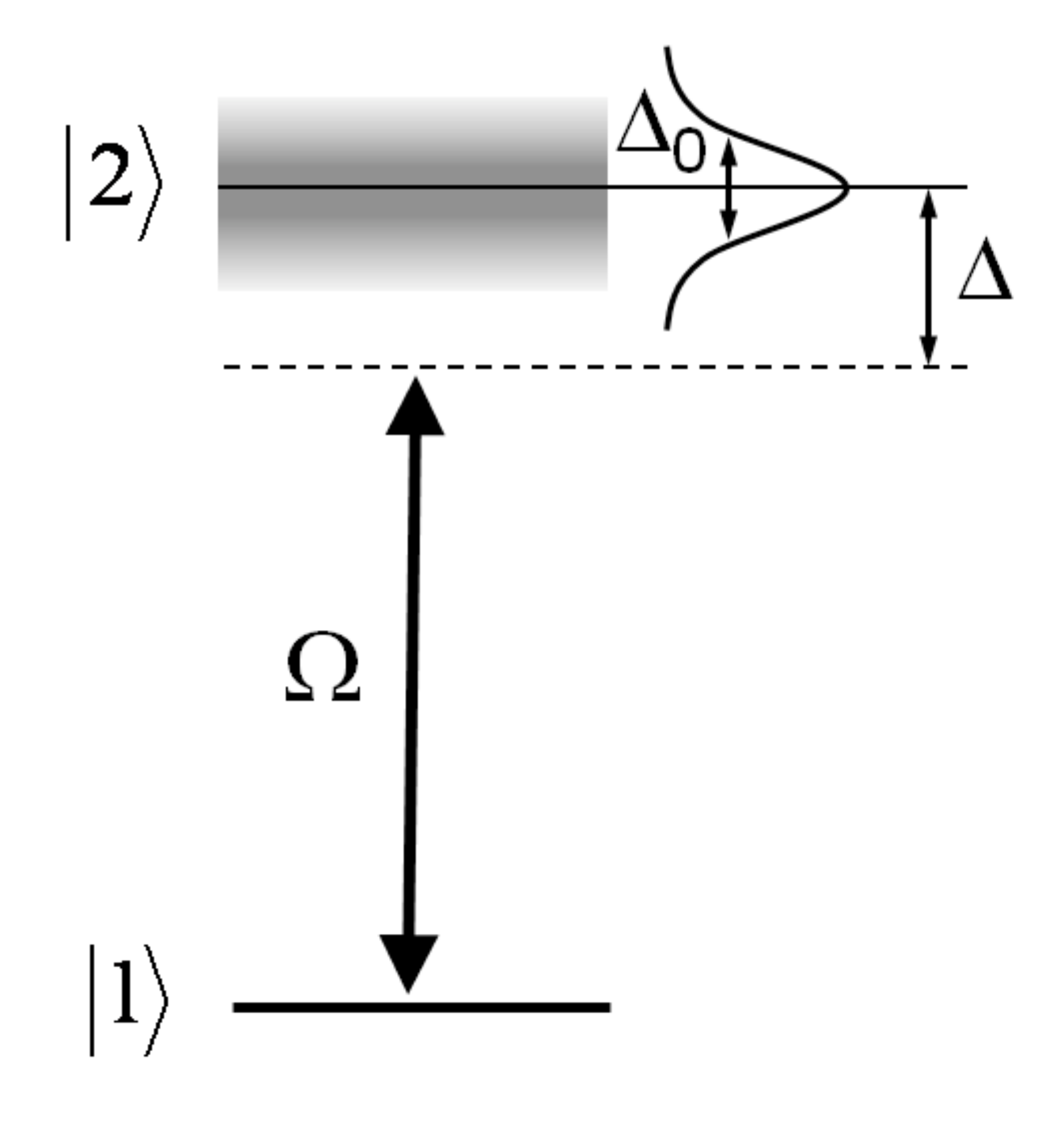}
\caption{Basic scheme of inhomogeneously broadened two-level systems with a ground state $|1\rangle$ and an excited state $|2\rangle.$ The spectral atomic distribution is supposed to have a Gaussian shape with the characteristic width $\Delta_0$. The interaction of the atomic ensemble with a laser pulse is parametrized by the Rabi frequency $\Omega$ and by the detuning $\Delta.$ }\label{fig1}}
\end{figure}

We start with an ensemble of two-level systems with a ground state $|1\rangle$ and an excited state $|2\rangle$ as depicted in Fig. \ref{fig1}. We suppose that, initially, all N atoms are prepared in the state $|1\rangle.$ We first look at the temporal evolution of a single atom which interacts with several laser pulses.\\

\textit{Response of a single atom} \\
Let us recall the expression of propagators associated to a two level atom under a pulsed excitation detuned from the resonance by $\Delta$ and with the Rabi frequency $\Omega_i(t).$ If the light pulse is short enough, i.e. its spectrum is much larger than the detuning $\Delta,$ the propagator is given by
\begin{eqnarray}
&& U^{\theta_i}(\tau)=\begin{pmatrix}
  \cos{\left(\theta_i/2\right)}    &  -i\sin{\left(\theta_i/2\right)}  \\
  -i\sin{\left(\theta_i/2\right)}    &  \cos{\left(\theta_i/2\right)}
\end{pmatrix},\\
&& \nonumber \theta_i=\int ds \Omega_i(s) \approx \Omega_i^{\text{max}} \tau
\end{eqnarray}
where $\tau$ is the temporal duration of the pulse and $\Omega_i^{\text{max}}$ is the maximum value of the Rabi frequency, i.e. $\Omega_i^{\text{max}}=\max_t \{\Omega_i(t)\}.$ Now consider the situation where the laser pulse is off. In this case, the propagator from time $t_i$ to $t_f$ reduces to
\begin{equation}
U^{\Delta}(t_f,t_i)=\begin{pmatrix}
  1    &  0  \\
  0    &  e^{-i\Delta (t_f-t_i)}
\end{pmatrix}.
\end{equation}

\begin{figure}[hr!]
{\includegraphics[scale=0.27]{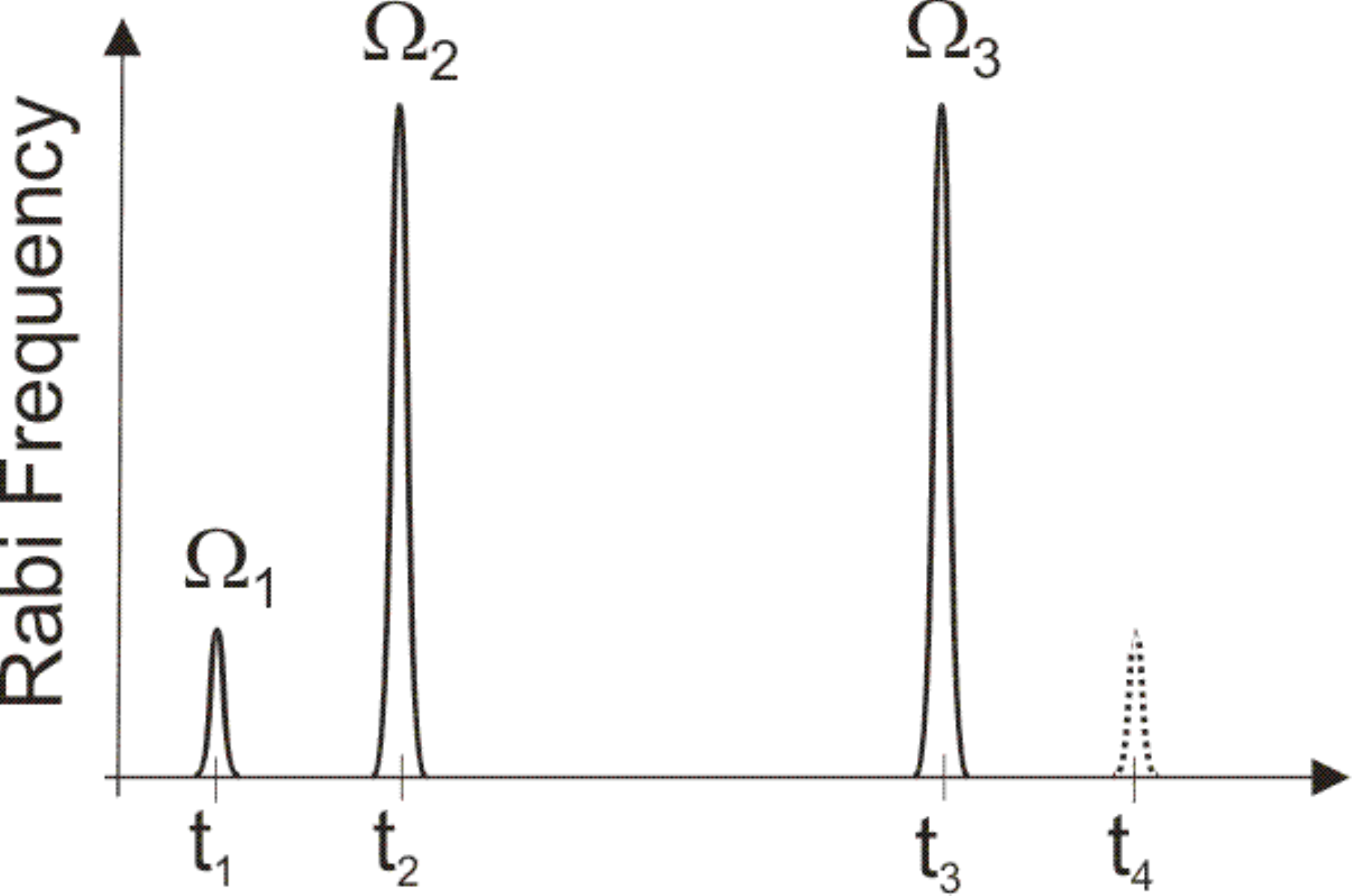}
\caption{Pulse sequence associated to the three-pulse photon echo in two-level atoms. The first pulse corresponding to $\Omega_1$ interacts with the atoms at time $t_1$ in order to be stored. The two next pulses, $\Omega_2$ and $\Omega_3$ applied at time $t_2$ and $t_3$ respectively, correspond to rephasing pulses. The echo signal is expected at time $t_4=t_3+t_2-t_1.$}\label{fig2}}
\end{figure}

Having this in mind, we calculate the state of the atom after a sequence of three pulses, cf. Fig. \ref{fig2}. We denote $a$ and $b$ the probability amplitudes of states $|1\rangle$ and $|2\rangle$ respectively. Since the atom is initially prepared in the ground state, we have $a(t_0=0)=1$ and $b(t_0=0)=0.$ The first pulse corresponding to the stored pulse and associated with the Rabi frequency $\Omega_1$ interacts with the atom at time $t_1.$ Then, two pulses are used to halt and to reverse the dephasing due to the inhomogeneous broadening. The interaction with the second pulse corresponding to $\Omega_2$ occurs at time $t_2.$ The probability amplitudes at $t_2$ are obtained by applying the following propagators $U^{\theta_2}(\tau)U^{\Delta}(t_2,t_1)U^{\theta_1}(\tau)U{^\Delta}(t_1,t_0)$ into the initial amplitudes, leading to
\begin{eqnarray}
\nonumber a(t_2)&=&\cos{\left(\theta_1/2\right)} \cos{\left(\theta_2/2\right)}\\
&-&\sin{\left(\theta_1/2\right)}\sin{\left(\theta_2/2\right)}e^{-i\Delta (t_2-t_1)},\\
\nonumber b(t_2)&=&-i\cos{\left(\theta_1/2\right)} \sin{\left(\theta_2/2\right)}\\
&-&i\sin{\left(\theta_1/2\right)}\cos{\left(\theta_2/2\right)}e^{-i\Delta (t_2-t_1)}.
\end{eqnarray}
We are interested in the case where the third pulse associated to $\Omega_3$ is applied at time $t_3$ such that $t_3-t_2$ is much longer than the phase relaxation time of the $|1\rangle$-$|2\rangle$ transition and much shorter than the radiative lifetime of the state $|2\rangle$. At time $t_3$ just before the interaction with the pulse $\Omega_3,$ the populations are identical to the ones at $t_2$ but the coherence between the states $|1\rangle$ and $|2\rangle$ is lost and the atomic state becomes
\begin{equation}
\rho(t_3)=|a(t_2)|^2 |1\rangle\langle 1|+|b(t_2)|^2 |2\rangle\langle 2|.
\end{equation}
At time $t_4>t_3,$ the atom is finally described by
\begin{eqnarray}
\nonumber
\rho(t_4)&=&|a(t_2)|^2 U^{\Delta}(t_4,t_3)U^{\theta_3}(\tau)|1\rangle\langle 1|U^{\theta_3\dagger}(\tau)U^{\Delta\dagger}(t_4,t_3)\\
\nonumber
&+&|b(t_2)|^2 U^{\Delta}(t_4,t_3)U^{\theta_3}(\tau)|2\rangle\langle 2|U^{\theta_3\dagger}(\tau)U^{\Delta\dagger}(t_4,t_3),
\end{eqnarray}
i.e. by a mixture of
\begin{eqnarray}
\nonumber
&U^{\Delta}(t_4,t_3)&U^{\theta_3}(\tau)|1\rangle=\\
&&\nonumber \cos\left(\theta_3/2\right)|1\rangle-i\sin\left(\theta_3/2\right)e^{-i\Delta (t_4-t_3)}|2\rangle
\end{eqnarray}
and
\begin{eqnarray}
\nonumber
&U^{\Delta}(t_4,t_3)&U^{\theta_3}(\tau)|2\rangle=\\
&&\nonumber-i\sin\left(\theta_3/2\right)|1\rangle+\cos\left(\theta_3/2\right)e^{-i\Delta (t_4-t_3)}|2\rangle
\end{eqnarray}
with weights $|a(t_2)|^2$ and $|b(t_2)|^2$ respectively.\\

\textit{Intensity of the echo signal:} \\
Consider the case where the first pulse corresponds to a few photon $\theta_1 \approx 2\epsilon.$ In this case, the number of atoms transfered into the state $|2\rangle$ is given by $N\sin(\theta_1/2)^2 \approx N\epsilon^2$ which is, in the single photon case, equal to $1$. When absorbed by the sample, this few-photon pulse induces a weak macroscopic polarization
\[
P(t_1)=\int_{-\infty}^{\infty} d\Delta g(\Delta) \-\ \wp \langle 1|\rho(t_1)|2\rangle
\]
\begin{equation}
\label{init_pola}
=\int_{-\infty}^{\infty} d\Delta g(\Delta) \-\ \wp \-\ a(t_1) b(t_1)^\star= i \-\ \wp \-\ N \epsilon
\end{equation}
where $\wp$ is the electric dipole moment and $g(\Delta)$ is the atomic spectral distribution, which satisfies $\int_{-\infty}^{\infty} d\Delta g(\Delta)=N.$ The light pulses in our model, in particular the pulse one wants to store, have spectral widths that are much larger than the width of the atomic distribution $g(\Delta)$, which we assume to have a Gaussian shape with the characteristic width $\Delta_0.$ The spectral band matching condition required for efficient absorption is thus not met. However, one can still bound the readout efficiency by comparing the initial polarization (\ref{init_pola}) and the one at the echo time, cf. below.

Within the context of three-pulse photon echo, the rephasing pulses that maximize the echo intensity are $\pi/2$ pulses, i.e. $\theta_2 =\theta_3 =\pi/2.$  The electric dipoles radiate in phase at time $t_4=t_3+t_2-t_1$ leading to a macroscopic polarization given by
\begin{eqnarray}
\nonumber
P(t_3+t_2-t_1)&=&\int_{-\infty}^{\infty} d\Delta g(\Delta) \wp \langle 1|\rho(t_4=t_3+t_2-t_1)|2\rangle\\
\nonumber
&=&-\int_{-\infty}^{\infty} d\Delta g(\Delta) \frac{i \wp \epsilon}{2}  (1+e^{2i\Delta (t_2-t_1)})\\
\label{final_pola}
&\rightarrow& \frac{-i\wp N \epsilon}{2} \-\ \text{for} \-\ (t_2-t_1)\gg 1/\Delta_0.
\end{eqnarray}
Hence, the initial polarization (see Eq. \ref{init_pola}) is not completely restituted. The efficiency of the storage protocol is thus intrinsically limited. Indeed, the polarization (\ref{final_pola}) of the atomic ensemble serves as a source for the collective emission of an echo that has an intensity given by
\begin{eqnarray}
\label{echo_2levels}
\nonumber I_{\text{echo}}&=&I_0 \Big|\int_{-\infty}^{\infty} d\Delta g(\Delta) \-\ \langle 1|\rho(t_4=t_3+t_2-t_1)|2\rangle\Big|^2\\
&=& \frac{1}{4}  N^2 \epsilon^2 I_0
\end{eqnarray}
where $I_0$ is the radiation intensity from a single isolated atom. The quantity $I_{\text{echo}}$ is to be compared with the intensity related to the absorbed part of the input pulse, which can be deduced from the initial polarization using
$$
I_{\text{1}}=I_0 \Big|\int_{-\infty}^{\infty} d\Delta g(\Delta) \-\ a(t_1)b(t_1)^\star\Big|^2
= N^2 \epsilon^2 I_0.
$$
The readout efficiency $I_{\text{echo}}/I_1$ (and thus the overall storage efficiency) is upper bounded by $1/4.$ Note that other detrimental effects due to the propagation within the atomic ensemble (free-induction decay resulting in the propagation of intense light pulses in the atomic medium or the finite value of the optical depth) further limit the efficiency \cite{Ruggiero09}. Let us now compare $I_{\text{echo}}$ with the incoherent emission from the same sample. \\

\textit{Fluorescence :} \\
To determine the amount of fluorescence, i.e. the spontaneous emission, we evaluate the population in the excited state at the echo time $t_3+t_2-t_1$ without having exposed the atoms to the first excitation, i.e. for $\epsilon=0.$ We easily obtain that
$
\langle 2 | \rho(t_3+t_2-t_1)|2\rangle = 1/2
$
meaning that the intensity associated to the spontaneous fluorescence is given by
 \begin{equation}
 \label{noise_2levels}
 I_{\text{noise}}= \frac{1}{2} N I_0.
 \end{equation}
The equations  (\ref{echo_2levels})-(\ref{noise_2levels}) remind us that the incoherent radiation as emanating from an ensemble of dipoles that are oscillating with random phases has an intensity proportional to $N$ whereas the collective emission coming from an ensemble of dipoles in phase gives rise to an echo with an intensity proportional to $N^2.$ \\
The ratio between the echo signal corresponding to a weak input and the noise is given by
\begin{equation}
\label{signaltonoise2}
\frac{I_{\text{echo}}}{I_{\text{noise}}}=\frac{1}{2}N\epsilon^2
\end{equation}
which reduces to $1/2$ in the particular case of a single-photon input where $N\epsilon^2=1$. This calculation shows that, due to the spontaneous emission that severely limits the achievable fidelity of the quantum state storage, the conventional three-pulse photon echo cannot be used to implement a quantum memory in two-level systems. Indeed, the fidelity for storage and recall of a time-bin qubit, i.e. a single photon in a superposition of two different emission times, $F=tr(\rho_{in}\rho_{out})$ with $\rho_{out}=(2F-1)\rho_{in}+(1-F)\opone$, is limited to $(1-F)=I_{noise}/(I_{echo}+2I_{noise})$, i.e. $F_{max}^{2-level}=3/5=0.6$. The value for the fidelity is smaller than the limit $F_{classical}=2/3$ imposed by optimum classical storage \cite{Massar1995,comment1}.\\

Note that this calculation concerns a sample of N atoms which is small compared with $\lambda^3.$
For larger samples, i.e. for sample sizes large compared with $\lambda^3,$ the spatial dependence of the electromagnetic fields must be taken into account. In this case, the polarization of the atomic ensemble at time $t_4=t_3+t_2-t_1$ for the specific mode associated to the wave vector ${\bf k}$ is found to be
\begin{eqnarray}
\nonumber
&&P(t_3+t_2-t_1,{\bf k})=-\int_{-\infty}^{\infty} d\Delta \-\  d{\bf r} \-\  g(\Delta,{\bf r}) \frac{i \wp \epsilon}{2} \\
\nonumber
&& \times  \left(e^{i({\bf k}+{\bf k_1}-{\bf k_2}-{\bf k_3}).{\bf r}}+e^{i({\bf k}-{\bf k_1}+{\bf k_2}-{\bf k_3}).{\bf r}}e^{2i\Delta (t_2-t_1)}\right).
\end{eqnarray}
${\bf k_i}$, $i=\{1,3\}$ refers the $i^{\text{th}}$ pulse and the atomic density now depends on the spatial coordinate ${\bf r}$. The exponential terms average to zero unless the phase matching condition is fulfilled ${\bf k} \approx -{\bf k_1}+{\bf k_2}+{\bf k_3}.$ When the sample is small compared with $\lambda^3$, the collective emission is isotropic and the signal-to-noise ratio (\ref{signaltonoise2}) holds for any direction. For sample sizes larger than $\lambda^3,$ the collective emission is highly directional. For the specific direction associated to the wave vector ${\bf k}$ satisfying the phase matching condition, the echo intensity is related to the spontaneous radiation intensity $I_0({\bf k})$ from a single atom by
$
I_{\text{echo}}({\bf k})=  \frac{1}{4}  N^2 \epsilon^2 I_0({\bf k}).
$
The noise intensity is given by
$
I_{\text{noise}}({\bf k})= \frac{1}{2} N I_0({\bf k})
$
and as before, the signal-to-noise ratio reduces to $1/2$ in the case of a single-photon storage. For another  direction, the signal field is obtained by squaring each individual dipole and then summing. This leads to a much worse signal-to-noise ratio of order $\epsilon^2.$ The conclusion is that a spatial filtering cannot be used to circumvent the problem of noise. Moreover, the simpler calculation where the spatial dependence of the fields is neglected, gives an upper bound for the signal-to-noise ratio. Now, we extend our analysis to three-level systems omitting the spatial dependence of the light.

\section{Three-level systems}
\label{section3}

\begin{figure}[hr!]
{\includegraphics[scale=0.27]{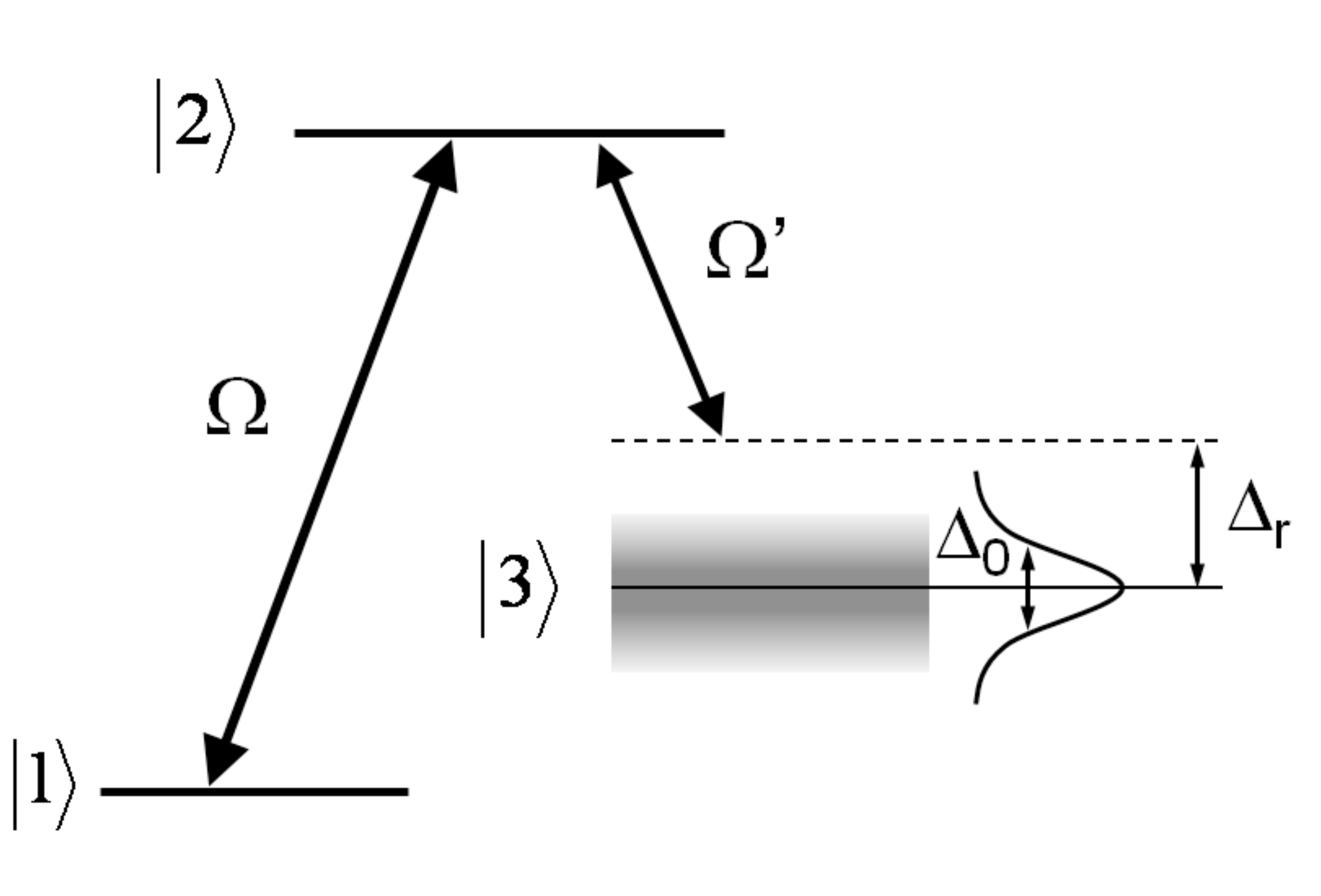}
\caption{Basic scheme of three-level systems with an excited state $|2\rangle$ and two ground states $|1\rangle$ and $|3\rangle$ that exhibits an inhomogeneous broadening on the spin transition. As in the two-level scheme, the atomic spectral distribution is supposed to be Gaussian with the characteristic width $\Delta_0$. The interaction of the atomic ensemble with a Raman laser pulse is parametrized by the effective Rabi frequency $\Omega^r_i=\sqrt{\Omega^2+\Omega'^2}$ and by the two-photon detuning $\Delta_r.$ }\label{fig3}}
\end{figure}

\begin{figure}[hr!]
{\includegraphics[scale=0.3]{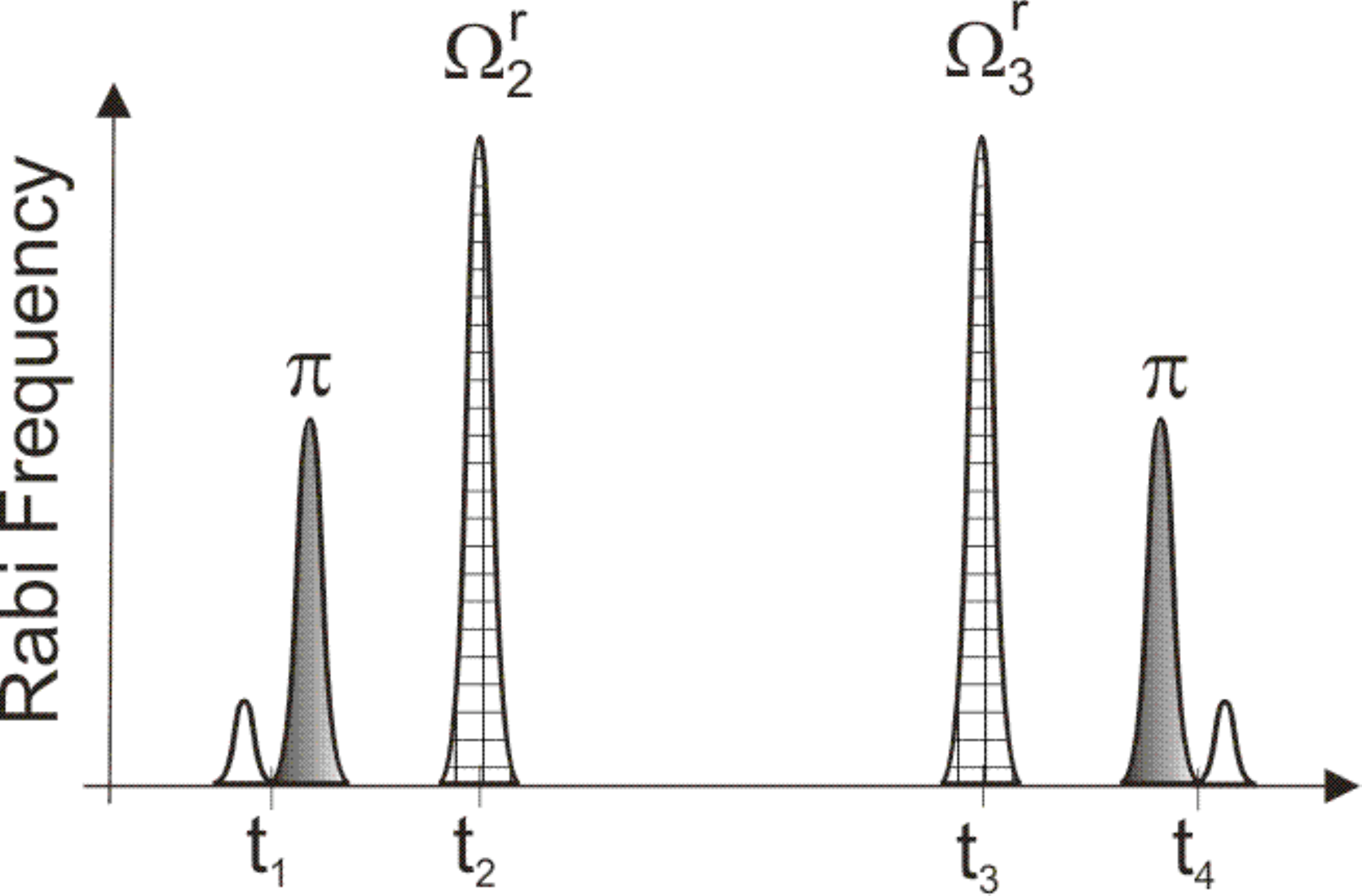}
\caption{Pulse sequence for the three-pulse photon echo in three-level systems. The first pulse, resonant with the transition $|1\rangle$-$|2\rangle$ is the one to store. It is immediately followed by a $\pi$-pulse at time $t_1$ on the transition $|2\rangle$-$|3\rangle$ to transfer the optical coherence into a spin coherence. The next two pulses, corresponding to $\Omega_2^r$ and $\Omega_3^r$, which are Raman pulses, force the ensemble of dipoles to rephase. The resulting coherences are read out at time $t_4=t_3+t_2-t_1$ by a $\pi$-pulse on the transition $|2\rangle$-$|3\rangle.$ The echo is emitted immediately after this final $\pi$ pulse.}\label{fig4}}
\end{figure}

A recent paper \cite{Ham09} suggests that the extension of the three-pulse photon echo to systems with additional ground states would allow to implement a quantum memory. In the following we apply the approach described above to quantify the fluorescence in three-level systems and we show that even in these systems, the photon-echo is inappropriate for the storage of single photons.\\

The basic scheme, motivated by the analysis of the proposal of Ref. \cite{Ham09}, is as follows. We start with an ensemble of three-level systems with an excited state $|2\rangle$ and two ground states $|1\rangle$ and $|3\rangle$. We consider that the optical transition $|1\rangle$-$|2\rangle$ is homogeneous and that the spin transition  $|1\rangle$-$|3\rangle$ exhibits an inhomogeneous broadening (see Fig. \ref{fig3}). As before, we first focus on the dynamics of a single atom interacting with a sequence of resonant pulses. To evaluate the response of the atomic ensemble, we sum over all individual responses to quantify the intensity of the echo and the fluorescence.\\

\textit{Response of a single atom :} \\
Consider a three-level atom under a pulsed Raman excitation corresponding to the effective Rabi frequency $\Omega^r_i(t)=\sqrt{\Omega^2_i(t)+\Omega'^2_i(t)}$ where $\Omega_i(t)$ ($\Omega'_i(t)$) is associated to the transition $|1\rangle$-$|2\rangle$ ($|2\rangle$-$|3\rangle$) and detuned from the two-photon resonance by $\Delta_r.$ Further suppose that the Rabi frequencies satisfied $\Omega^{\max}_i \approx \Omega'^{\max}_i$ and that the pulse durations $\tau$ are short enough so that $\tau\ll 1/\Delta_0,$ as before. In the basis $\{|1\rangle, |2\rangle, |3\rangle\},$ the propagator is given by
\begin{eqnarray}
\nonumber &U^{\theta^r_i}(\tau)&=\begin{pmatrix}
  \frac{1}{2}\left(1+\cos\frac{\theta^r_i}{2}\right)    &   \frac{-i}{\sqrt{2}}\sin\frac{\theta^r_i}{2}  &   \frac{1}{2}\left(-1+\cos\frac{\theta^r_i}{2}\right) \\
   \frac{-i}{\sqrt{2}}\sin\frac{\theta^r_i}{2}   &   \cos\frac{\theta_i^r}{2}   &   \frac{-i}{\sqrt{2}}\sin\frac{\theta^r_i}{2}\\
   \frac{1}{2}\left(-1+\cos\frac{\theta^r_i}{2}\right)   & \frac{-i}{\sqrt{2}}\sin\frac{\theta^r_i}{2} &   \frac{1}{2}\left(1+\cos\frac{\theta^r_i}{2}\right)     \\
\end{pmatrix}\\
\nonumber 
\end{eqnarray}
\noindent
with $\theta_i^r\approx\Omega^{r,\text{max}}_i \tau.$ If the laser pulses are off, the propagator reduces to
\begin{equation}
U^{\Delta_r}(t_f,t_i)=\begin{pmatrix}
  1    &  0 &  0  \\
  0    &  1 &  0\\
  0 &  0 &  e^{-i\Delta_r (t_f-t_i)}
\end{pmatrix}.
\end{equation}

A possible procedure to transfer properly a weak light excitation into a spin coherence requires first the preparation of atoms in one of the two ground states, say, the state $|1\rangle.$ The weak pulse that one wants to store is resonant with the transition $|1\rangle$-$|2\rangle$ and generates a weak coherence $i\epsilon$ between the states $|1\rangle$ and $|2\rangle.$ By applying immediately a $\pi$-pulse on the transition $|2\rangle$-$|3\rangle,$ the optical coherence is transfered to a coherence between the state $|1\rangle$ and the state $|3\rangle.$ At time $t_1$ after this sequence of pulses, the probability amplitudes $a,$ $b$ and $c$ associated to the states $|1\rangle,$ $|2\rangle$ and $|3\rangle$, respectively, are given by
\begin{equation}
\label{coherence_weak}
a(t_1)=1, b(t_1)=0, c(t_1)=-\epsilon,
\end{equation}
i.e. the weak pulse information is correctly stored into a spin coherence. \\

Note at this stage that if one would use rephasing pulses that are off-resonant with the $|1\rangle$-$|2\rangle$ and $|2\rangle$-$|3\rangle$ transitions, the protocol would be identical with the above-discussed three-pulse echo in two-level systems. Non-resonant light could indeed compensate the dephasing associated with the inhomogeneously broadened spin transition without population transfer to the excited state $|2\rangle$, which could thus be eliminated adiabatically. The two spin states $|1\rangle$ and $|3\rangle$ would be equally populated at the rephasing time. As a consequence, the last $\pi$-pulse used to transfer the spin coherence to an optical coherence, would transfer the population of one of ground states into the excited state. The resulting spontaneous emission would lead to a signal-to-noise ratio equal to the one given in Eq. (\ref{signaltonoise2}).\\

But let us consider resonant rephasing light pulses applied simultaneously on the transitions $|1\rangle$-$|2\rangle$ and $|2\rangle$-$|3\rangle$, as in Ref. \cite{Ham09}. More precisely, two Raman pulses corresponding to $\theta_2^r$ and $\theta_3^r$ are sent through the medium at time $t_2$ and $t_3$, respectively, to compensate the inhomogeneous dephasing. Further, consider the case where the delay $t_3-t_2$ between these Raman pulses is much longer than the phase relaxation time of the $|1\rangle$-$|3\rangle$ transition, as before. Finally, at time $t_4>t_3$, a $\pi$-pulse resonant with the transition $|2\rangle$-$|3\rangle$ transfers back the spin coherence into an optical coherence. At time $t_4$ just after the interaction with this last $\pi$-pulse, the state of the atom is described by a mixture of three states
\begin{eqnarray}
\nonumber |\psi_1\rangle&=& \frac{1}{2}\left(1+\cos\frac{\theta_3^r}{2}\right)|1\rangle \\
&&\nonumber -\frac{i}{2}\left(-1+\cos\frac{\theta_3^r}{2}\right)e^{-i\Delta_r (t_4-t_3)}|2\rangle-\frac{1}{\sqrt{2}}\sin\frac{\theta_3^r}{2}|3\rangle,\\
\nonumber |\psi_2\rangle&=&  -\frac{i}{\sqrt{2}}\sin\frac{\theta_3^r}{2}|1\rangle\\
&&\nonumber -\frac{1}{\sqrt{2}}\sin\frac{\theta_3^r}{2}e^{-i\Delta_r (t_4-t_3)}|2\rangle-i\cos\frac{\theta_3^r}{2}|3\rangle,\\
\nonumber |\psi_3\rangle&=& \frac{1}{2}\left(-1+\cos\frac{\theta_3^r}{2}\right)|1\rangle \\
&&\nonumber -\frac{i}{2}\left(1+\cos\frac{\theta_3^r}{2}\right)e^{-i\Delta_r (t_4-t_3)}|2\rangle-\frac{1}{\sqrt{2}}\sin\frac{\theta_3^r}{2}|3\rangle,
\end{eqnarray}
with respective weights $|a(t_2)|^2,$ $|b(t_2)|^2$ and $|c(t_2)|^2$ given by
\begin{eqnarray}
\nonumber &a(t_2)&=\frac{1}{2}\left(1+\cos\frac{\theta_2^r}{2}\right)-\frac{\epsilon}{2}\left(-1+\cos\frac{\theta_2^r}{2}\right)e^{-i\Delta_r (t_2-t_1)}, \\
\nonumber &b(t_2)&=-\frac{i}{\sqrt{2}}\sin\frac{\theta_2^r}{2}+\frac{i\epsilon}{\sqrt{2}}\sin\frac{\theta_2^r}{2}e^{-i\Delta_r (t_2-t_1)},\\
\nonumber &c(t_2)&=\frac{1}{2}\left(-1+\cos\frac{\theta_2^r}{2}\right)-\frac{\epsilon}{2}\left(1+\cos\frac{\theta_2^r}{2}\right)e^{-i\Delta_r (t_2-t_1)}.
\end{eqnarray}

\textit{Intensity of the echo signal:} \\
When absorbed by the sample, the first weak pulse induces a weak macroscopic polarization
\begin{equation}
\label{initial_pola}
P(t_1)= i \wp N \epsilon.
\end{equation}
If the two Raman pulses are $\pi$-Raman pulses, i.e. $\theta_2^r =\theta_3^r =\pi$ as suggested in Ref. \cite{Ham09}, they lead to a macroscopic polarization at time $t_4=t_3+t_2-t_1$ given by
\begin{eqnarray}
\nonumber
P(t_4=t_3+t_2-t_1)&\rightarrow& \frac{-i3\wp  N \epsilon}{8}  \-\ \text{for} \-\ (t_2-t_1)\gg 1/\Delta_0.
\end{eqnarray}
As in two-level systems, the initial polarization (\ref{initial_pola}) is not completely restituted, leading to a fundamental limitation for the storage efficiency. The intensity of the echo signal is related to the intensity associated with the spontaneous emission coming from a single atom $I_0$ by
\begin{equation}
I_{\text{echo}}= \frac{9}{64} N^2 \epsilon^2 I_0.
\end{equation}
Since the intensity related to the absorbed part of the input pulse is equal to
$
N^2 \epsilon^2 I_0,
$
the storage efficiency is upper bounded by $9/64$ (roughly 15\%).\\

\textit{Fluorescence :} \\
To determine the amount of noise, we evaluate the population in the excited state at the echo time $t_4=t_3+t_2-t_1$ without excitation, i.e. for $\epsilon=0,$ as in two-level systems. We find
\begin{equation}
I_{\text{noise}} = \frac{3}{8} N I_0
\end{equation}
so that the signal-to-noise ratio is given by
\begin{equation}
\frac{I_{\text{echo}}}{I_{\text{noise}}} = \frac{3}{8}N\epsilon^2,
\end{equation}
which reduces to $3/8$ in the single-photon case corresponding to $N\epsilon^2=1.$ Again, due to the spontaneous emission induced by the rephasing pulses, the fidelity of the quantum state storage is degraded beyond the classical limit: $F_{max}^{3-level}=\frac{11}{19}\approx 0.58$. Note that $F_{max}^{3-level}<F_{max}^{2-level}$.\\

Before concluding, let us focus on the analysis presented in Ref. \cite{Ham09}. The author uses four-level systems with three ground states $|1\rangle,$ $|3\rangle,$ $|a\rangle$ and one excited state $|2\rangle$. The studied protocol is similar to the above discussed one, except that the excited atoms are reversibly transfered to the auxiliary state $|a\rangle$ during storage, i.e. between the times $t_2$ and $t_3.$ This population transfer to $|a\rangle$ avoids the loss by spontaneous emission between times $t_2$ and $t_3.$ However, the upper level population is restored at time $t_3,$ before applying the second Raman $\pi$ pulse. Hence, one exactly recovers the situation we considered above. As a consequence, the problem of the spontaneous emission is not circumvented and the signal is equally dominated by the noise.

\section{Conclusion}
In this paper, we proved that the conventional three-pulse photon echo cannot be used for quantum storage. A fundamental limitation is related to the medium excitation induced by the rephasing pulses.
This excitation induces spontaneous emission, which produces a noise comparable to the retrieved signal and thus deteriorates the storage fidelity beyond the classical limit. This is a major difference compared to genuine quantum memory protocols such as stopped light based on electromagnetically induced transparency (EIT) \cite{Fleisch00}, controlled reversible inhomogeneous broadening (CRIB) \cite{Nilson05, Alexander06, Kraus06, Sangouard07}, or atomic frequency combs (AFC) \cite{Afzelius09}, where the control pulses transfer a negligible number of atoms to excited states.
Note finally that additional detrimental effects due to propagation \cite{Ruggiero09} could further limit the fidelity and the efficiency of quantum storage based on three-pulse photon echo.

\section*{Acknowledgements}
The Geneva authors acknowledge financial supports from NCCR Quantum Photonics and from the EU project QuReP. W.T. acknowledges funding by NSERC, iCORE and GDC.


\begin{thebibliography}{100}

\bibitem{Sangouard09}  N. Sangouard, C. Simon, H. De Riedmatten, and N. Gisin, arXiv:0906.2699; C. Simon \textit{et al.}, Phys. Rev. Lett. \textbf{98}, 190503 (2007).

\bibitem{Mossberg82} T.W. Mossberg, Opt. Lett. \textbf{7}, 77 (1982).

\bibitem{Iin95} H. Lin, T. Wang, and T.W. Mossberg, Opt. Lett. \textbf{20}, 1658 (1995).

\bibitem{Ohlsson03} N. Ohlsson, M. Nilsson, S. Kroll, and R.K. Mohan, Opt. Lett. 28, 450-452 (2003)

\bibitem{Ruggiero09} J. Ruggiero, J.-L. Le Gou\"et, C. Simon, and T. Channelière, Phys, Rev. A \textbf{79},053851 (2009).

\bibitem{Ham09} B.S. Ham, Nat. Photonics 3, 518 (2009).

\bibitem{Abella66} I.D. Abella, N.A. Kurnit, and S.R. Hartmann, Phys. Rev. \textbf{141}, 391 (1966).

\bibitem{Mossberg79} T.W. Mossberg, R. Kachru, and S.R. Hartmann, Phys. Rev. A \textbf{20}, 1976 (1979).


\bibitem{Massar1995} S. Massar and S. Popescu, Phys. Rev. Lett. \textbf{74}, 1259 (1995).

\bibitem{comment1} It is interesting to note that the classical limit is saturated using a two-pulse photon-echo storage approach \cite{Ruggiero09}. The decay of the excited coherences in the here discussed, three-pulse echo approach is at the origin of the additionally reduced fidelity.

\bibitem{Fleisch00} M. Fleischhauer and M.D. Lukin, Phys. Rev. Lett.  \textbf{84}, 5094 (2000).

\bibitem{Nilson05} M. Nilson and S. Kr\"oll, Opt. Commun. \textbf{247}, 393 (2005).

\bibitem{Alexander06} A.L. Alexander \textit{et al.}, Phys. Rev. Lett. \textbf{96}, 043602 (2006).

\bibitem{Kraus06} B. Kraus \textit{et al.}, Phys. Rev. A. \textbf{73}, 020302(R) (2006).

\bibitem{Sangouard07} N. Sangouard, C. Simon, M. Afzelius, and N. Gisin, Phys. Rev. A. \textbf{75}, 032327 (2007).

\bibitem{Afzelius09} M. Afzelius, C. Simon, H. De Riedmatten, and N. Gisin, Phys. Rev. A. \textbf{79}, 052329 (2009).



\end{thebibliography}
\end{document}